\documentstyle[aps,pra,amsmath,amssymb,citesort,epsfig,multicol]{revtex}
\begin{document}

\title{Interfacial tension and nucleation in mixtures of
colloids and long ideal polymer coils}

\author{{\bf Richard P. Sear}\\
~\\
Department of Physics, University of Surrey,\\
Guildford, Surrey GU2 7XH, United Kingdom\\
email: r.sear@surrey.ac.uk}

\maketitle

\begin{abstract}
Mixtures of ideal polymers with hard spheres whose diameters are smaller than
the radius of gyration of the polymer, exhibit extensive immiscibility.
The interfacial tension between demixed phases of these
mixtures is estimated, as
is the barrier to nucleation. The
barrier is found to scale linearly with the radius of the polymer,
causing it to become large for large polymers. Thus for large polymers
nucleation is suppressed and phase separation proceeds via
spinodal decomposition, as it does in polymer blends.
\end{abstract}

\begin{multicols}{2}

\section{Introduction}

In earlier work \cite{sear01} immiscibility in
mixtures of colloidal hard spheres
and long ideal polymer molecules was studied.
Long means that the root-mean-square
end-to-end separation of the polymer molecules, $R_E$, is larger
than the diameter $\sigma$ of the hard spheres.
A mixture of spheres and long polymers was found to demix at
comparable number densities
of polymer molecules and spheres, both densities
scale with $R_E$ as $1/R_E^2\sigma$, for $R_E>\sigma$. This
scaling comes directly from the leading order term in the
second virial coefficient for the sphere-polymer interaction. The
interaction and hence the
virial coefficient must be extensive in the number of monomers for large $R_E$
and hence must scale as $R_E^2$ for our ideal polymers. The requirement
that it has the dimensions of a volume then imposes the scaling
$R_E^2\sigma$ as $\sigma$ is the only other relevant length in the
problem: the monomer size is assumed to be much less than $\sigma$
and so is irrelevant.
Once the mixture has demixed we have two coexisting phases:
one with a high density of colloidal particles and a low density
of polymer molecules, and one with a high density of polymer molecules
and a low density of colloidal particles. There is an
interface between these two coexisting phases. Here we determine the
scaling of the interfacial tension $\gamma$ of this interface,
and use it to show that when a mixed sample of polymer and colloid
is prepared and quenched into the two-phase region, the dynamics
of the separation into two phases starts off with spinodal
decomposition not nucleation. The fact that the
phase separation starts off with spinodal decomposition
makes mixtures of the
long polymers like polymer blends but unlike simple mixtures, e.g.,
mixtures of oil and water. Thus, we can apply much of what
we have learned of spinodal decomposition in systems like polymer
blends, to mixtures of colloidal spheres and much larger polymer molecules.

The colloid-polymer interactions in and bulk thermodynamics of,
mixtures of colloids and large ideal polymers have both been studied,
see
Refs.~
\cite{eisenriegler96,hanke99,meijer94,sear01,chatterjee98,odijk00,tuinier00},
but this is the first study of the interfacial tension and nucleation
of these systems.
The opposite limit to that of interest here, i.e., where the
polymer molecules are smaller than the colloidal spheres, has
been considered extensively, see
Refs.~\cite{gast83,lekkerkerker92,meijer94,dijkstra99}
for work on the bulk phase behaviour and Refs.~\cite{vrij97,brader00,evans01}
for work on interfaces. The following two sections deal
with the interfacial tension, and with nucleation. Throughout, the
objective will be to determine the scaling of the behaviour with
the ratio of the size of the polymer to that of the sphere.
Also note that here the polymers are always ideal, mixtures
of polymers with strong excluded volume interactions and spheres,
are very different \cite{maassen01,bolhuis01,fuchs02,tobe}.

\section{Interfacial tension between the demixed phases}

The interfacial tension between the demixed phases, one colloid-rich,
the other polymer rich, can be estimated using just dimensional
analysis. The tension $\gamma$ is an energy per unit area.
It is obtained by multiplying the free energy per unit volume,
which is $kT/(R_E^2\sigma)$ \cite{sear01}, by the width of the
interface. This width will be of the order of the polymer size $R_E$.
Thus, $\gamma\sim kT/R_E\sigma$. Note that the energy scale
has to be the thermal energy $kT$ as there are no other relevant energy
scales in the problem. The mixture is athermal, there
are no attractive interactions or soft repulsions to provide another
energy scale. The free-energy density is then of order $kT$ times
the number density, which is of order $1/R_E^2\sigma$ for both
the polymer molecules and the colloidal spheres when they demix.
This is just a simple scaling argument so we confirm it by
determining the scaling of $\gamma$ within a standard square-gradient
or Cahn-Hilliard theory for the interface
\cite{cahn58,evans79,chaikin,binder87,debenedetti}.

We apply this theory to the system in the semigrand ensemble
of Ref.~\cite{sear01} where the characteristic thermodynamic
potential is the semigrand potential $\omega$, which is a function
of the number density of colloidal particles, $\rho_C$, and
the activity $z$ of the polymer molecules. As we are specifying
the activity not the density of the polymer our system is equivalent
to a single component system whose thermodynamic
state depends on the density and on the activity of the
polymer $z$; $\ln z$ acts as an inverse temperature in the sense
that the larger it is the stronger is the effect of the attractions.
Thus we can apply the standard
square-gradient expression for the interfacial tension of a single
component system, which is \cite{cahn58,evans79,debenedetti,chaikin,brader00}
\begin{equation}
\gamma=\int{\rm d}x\left[\Psi+
\kappa\left(\frac{{\rm d}\rho_C}{{\rm d}x}\right)^2\right],
\label{sq}
\end{equation}
where
\begin{equation}
\Psi=\omega(\rho_C(x))-\omega^{(b)}-\mu_C(\rho_C(x)-\rho_C^{(b)}),
\end{equation}
is the excess grand potential at a point.
$\omega^{(\alpha)}$ and $\rho_C^{(\alpha)}$ are the semigrand potential
and density in either one of the coexisting phases. The superscript
$\alpha=C,P$ for the colloid-rich and polymer-rich phases respectively.
$\mu_C$ is the
chemical potential of the colloid. The interface is normal to the
$x$-axis. The coefficient $\kappa$ of the gradient term is
assumed to be density independent. The equilibrium profile is obtained
by minimising Eq.~(\ref{sq}). Then standard manipulations enable
a simpler expression for the equilibrium interfacial tension
to be derived \cite{cahn58}
\begin{equation}
\gamma=2\int_{\rho_C^{(P)}}^{\rho_C^{(C)}}
{\rm d}\rho_C\left[\kappa\Psi\right]^{1/2}.
\label{sq_eq}
\end{equation}

If we require that the functional Eq.~(\ref{sq}) be
consistent with linear response theory \cite{evans79} we obtain
an expression for the coefficient $\kappa$ of the gradient term
\begin{equation}
\kappa=\frac{kT}{12}\int{\rm d}{\bf r}r^2c_2(r;\rho_C,z),
\label{kappa}
\end{equation}
where $c_2(r;\rho_C,z)$ is the direct correlation function of the fluid of
colloidal hard spheres in the presence of polymer. For our
systems the most basic assumption is to use the low density
approximation to the direct correlation function. This replaces $c$
with the Mayer f function for the effective sphere-sphere interaction
in the presence of the polymer \cite{evans79}. For two spheres with centres
separated by less than $\sigma$, the interaction energy is infinite
and the Mayer f function equals $-1$. For separations larger than
$\sigma$ the only interaction is that due to the polymer. This interaction
is known \cite{hanke99},
and is long-range and weak thus we linearise the Mayer f function. Adding this
altogether we obtain
\begin{equation}
c(r;\rho_C,z)\sim\left\{\begin{array}{cc}
-1 & r<\sigma \\
zR_E^2\sigma\left(\sigma/r\right) & \sigma<r\lesssim R_E \\
0 & r\gg R_E \\
\end{array},\right.
\end{equation}
the ideal polymer induces an attraction which decays as $1/r$ for
separations less than the radius of the polymer and
roughly exponentially beyond this.
Putting our approximate $c$ into Eq.~(\ref{kappa}) we obtain
an estimate of this coefficient
\begin{eqnarray}
\kappa &\sim & kTzR_E^2\sigma^2\int_0^{R_E}{\rm d}rr^3\\
&\sim& kTz\sigma^2R_E^6.
\label{kscale}
\end{eqnarray}

We now return to Eq.~(\ref{sq_eq}) for the interfacial tension
and determine its scaling with $R_E$. We note that the
density difference $\rho_C^{(C)}-\rho_C^{(P)}\sim 1/R_E^2\sigma$,
$\kappa$ scales as given by Eq.~(\ref{kscale}),
$\Psi\sim kT/R_E^2\sigma$ and the polymer activity is
of order the polymer number density $z\sim 1/R_E^2\sigma$.
Putting this all together we see that
we recover the scaling $\gamma\sim kT/R_E\sigma$ obtained earlier
by dimensional analysis. Also, from Eq.~(\ref{sq}) we see that the
characteristic length-scale for the interface must be
$(\kappa/\Psi)^{1/2}\rho_C^{(\alpha)}\sim R_E$, as we assumed earlier.
Earlier work by Vrij \cite{vrij97}, and by
Brader and Evans \cite{brader00}
on the interfacial tension between
demixed colloid-rich and polymer-rich phases when the colloid and
polymer were of comparable sizes, $R_E\sim\sigma$, found that,
as expected. $\gamma\sim kT/\sigma^2\sim kT/R_E^2$.
This is consistent with experimental findings \cite{hoog99}.

The interfacial tension $\gamma$ will be of order $kT/R_E\sigma$ only
if we are not too close to the critical point of the polymer-colloid
demixing. In general we have $\gamma=(kT/R_E\sigma)s(z/z_c-1)$,
where $s$ is a scaling function and $z_c$ is the polymer activity
at the critical point. We have been assuming that we are not very
close to the critical point, i.e., that $z/z_c-1=O(1)$, and for these values of
its argument $s=O(1)$ and we return to $\gamma$ being of order
$kT/R_E\sigma$. However, as the critical point is approached,
$z/z_c-1\ll 1$, we have that
the scaling function $s(x)=s_0x^{\mu}$ for $x\ll1$, where $s_0$
is a dimensionless constant and $\mu$ is the (positive) critical exponent
of the interfacial tension \cite{widom85}. The
interfacial tension tends to 0 as the critical
point of demixing is approached, and near the critical point it
varies as a power law.
See the review of Widom \cite{widom85}
for an excellent introduction to interfaces near critical points.
Sufficiently close to the critical
point the scaling of the interfacial tension will be dominated
by fluctuations and then the exponent $\mu$ will take the value
for the Ising model in three dimensions, $\mu=1.26$ \cite{widom85,chaikin}.
However, for very long polymers $R_E\gg\sigma$ the effective interaction
is long-ranged and long-range interactions suppress fluctuations and make
the system mean-field--like.
The mean-field value of the exponent $\mu$ is $3/2$ \cite{widom85}.
Which value of the exponent, Ising or mean-field, is observed is
determined by whether or not the Ginzburg criterion is
obeyed or not; see Ref.~\cite{chaikin} or any introduction
to critical phenomena for a definition of the Ginzburg criterion.
Note that Eq.~(\ref{sq_eq}), belonging as it does to a mean-field theory,
will yield an interfacial tension which tends to 0 with an exponent
$\mu=3/2$, its mean-field value.

\section{Nucleation and other fluctuations}

Now consider a single phase mixture of spheres and polymer
quenched into the two-phase coexistence region. For definiteness
assume that the single phase is the polymer-rich one. Then in order for the
second, colloid-rich, phase to form and coexist with the polymer-rich one,
this second phase must form.
The dynamics of the formation of a new phase fall into two broad
categories: nucleation then growth, and spinodal decomposition.
See Refs.~\cite{debenedetti,binder87,chaikin} for an introduction
to the dynamics of first-order phase transitions.
For example a mixture of
simple liquids such as water and an alcohol phase separate
via nucleation of the new water-rich or alcohol-rich phase followed
by growth of the nuclei, whereas polymer blends phase separate
via spinodal decomposition. Here we show that for large ideal polymers
and spheres,
nucleation becomes very difficult, so mixtures of large ideal
polymers and much smaller spheres will start to phase separate
via spinodal decomposition.

The rate of nucleation $N_n$ can be estimated using classical
nucleation theory,
see the book of Debenedetti \cite{debenedetti} for a
comprehensive discussion, see also Refs.~\cite{chaikin,binder87}.
$N_n$ is the number of nuclei crossing the barrier per unit time
per unit volume.
The classical nucleation theory expression for the rate $N_n$ is
\begin{equation}
N_n=\Gamma\exp(-\Delta F^*/kT),
\label{cnt}
\end{equation}
where $\Gamma$ is an attempt frequency per unit volume, generally
slowly varying, and $\Delta F^*$ is the free energy barrier that
must be crossed in order for a new phase to nucleate. The
variation in the rate is generally dominated by that in $\Delta F^*$
so we focus on this. The free energy barrier comes from the free energy
needed to form a microscopic droplet of the new phase, here a colloid-rich
phase. This droplet is the nucleus of the new phase.
Within classical nucleation theory the free energy of formation
of a microscopic droplet is the sum of two terms, a bulk term and a surface
term,
\begin{equation}
\Delta F=\frac{4}{3}\pi R^3\Psi_n+4\pi R^2\gamma,
\label{df1}
\end{equation}
where $R$ is the radius of the droplet and
\begin{equation}
\Psi_n=\omega(\rho_C^{(n)})-\omega(\rho_C)-\mu_C(\rho_C^{(n)}-\rho_C)
\label{psin1}
\end{equation}
is the difference between the grand potential
inside the nucleus and the grand potential of the
phase in which the nucleus forms,
$\rho_C$ is the density of colloid in the phase in which the nucleus forms,
and $\rho_C^{(n)}$ is the density of colloid inside the nucleus.
So long as we do not approach the spinodal too closely, we
can express $\Psi_n$ as a Taylor
expansion in chemical potential,
around the chemical potential of the colloid at
coexistence $\mu_{co}$.
Truncating the Taylor series after the linear term, we get
\begin{equation}
\Psi_n\simeq \Psi'\left(\mu_C-\mu_{co}\right),
\label{psin2}
\end{equation}
as $\Psi_n=0$ at coexistence, with
\begin{equation}
\Psi'=\left(\frac{\partial \Psi_n}
{\partial\mu_C}\right)_{\mu_C=\mu_{co}},
\label{psid}
\end{equation}
the derivative of $\Psi_n$ at coexistence.

The barrier is given by the free energy of the droplet whose free
energy is highest, which occurs when the two terms in Eq.~(\ref{df1})
are comparable,
$R^3\Psi_n\sim R^2\gamma$. Thus,
$R\sim \gamma/(\Psi'(\mu_C-\mu_{co}))$, and
\begin{equation}
\Delta F^* \sim\frac{\gamma^3}
{\left(\Psi'\left(\mu_C-\mu_{co}\right)\right)^2}.
\label{df_s1}
\end{equation}
From Eq.~(\ref{psid}),
$\Psi'$ scales as $1/R_E^2\sigma$. Using this scaling together with
that of $\gamma$, we find that at the top of the barrier
the size of the nucleus
is of order $R_E$, for $\mu_C-\mu_{co}$ not too much less than $kT$.
Using these same scalings in Eq.~(\ref{df_s1}), we obtain the
principle result of this section, the scaling of
the nucleation barrier $\Delta F^*$,
\begin{equation}
\frac{\Delta F^*}{kT} \sim\left(\frac{\mu_{co}}
{\mu_C-\mu_{co}}\right)^2
\frac{R_E}{\sigma},
\label{df_scale}
\end{equation}
the barrier scales as $R_E/\sigma$ and so increases as the
size of the polymer molecules relative to that of the colloidal
spheres increases. The factor in parentheses is a
dimensionless measure of the supersaturation: how far we are into the
two-phase region.

For sufficiently large $R_E$ of the
polymers the nucleation barrier will become so large that
nucleation cannot occur. The mixture will be metastable up to very
close to the spinodal \cite{herrmann82,binder83,binder84,binder87}.
Thus, the mixture will only start to
demix when quenched beyond the spinodal, where the phase separation
will start with spinodal decomposition. This is precisely analogous
to polymer blends and systems of particles in which the particle-particle
attractions are long ranged. In these systems the nucleation
barrier scales as $N^{1/2}$ and as $r^3$, where $N$ is the length
of the polymer and $r$ is the range of the attraction \cite{binder84}.
The phase
transition dynamics of systems of polymers and of particles with long-range
attractions, were studied extensively in the early 1980s by Binder and
Klein and their coworkers \cite{herrmann82,binder83,binder84}.
Many of the conclusions of that work also apply to the demixing
of mixtures of hard spheres and much larger ideal polymers.

Finally, we note that our finding that nucleation is suppressed is
equivalent to saying that our mixture satisfies the Ginzburg
criterion for the irrelevance of fluctuations \cite{chaikin}.
Essentially, nucleation {\em is} a fluctuation so when fluctuations
are weak nucleation is suppressed and vice versa, again
this was found for polymer blends/particles with long-range
attractions \cite{degennes,binder84}. For our mixtures,
the Ginzburg criterion is essentially
that the root-mean-square (rms) fluctuations in the number of
colloidal spheres or of polymer molecules, in
a volume $R_E^3$, are much less than the mean number in that volume.
The volume $R_E^3$ is the volume over which a pair of spheres or
polymer interact (the sphere-sphere interaction is mediated
by the polymer molecules, and the polymer-polymer interaction
is mediated by the spheres).
The rms fluctuations scale as the square root of the
number of spheres in a volume $R_E^3$, which is $\sim (R_E/\sigma)^{1/2}$,
whereas the mean number scales as $R_E/\sigma$. Thus for large
values of the ratio of the sizes, $R_E/\sigma$, the rms fluctuations
in the number
of spheres (or polymer molecules) inside the interaction volume is a small
fraction of the mean number. Fluctuations about the mean are small
and so mean-field theory applies and nucleation, which is a
fluctuation, has a very high free-energy cost.


\end{multicols}

\end{document}